\documentclass[conference, letterpaper]{IEEEtran}

\IEEEoverridecommandlockouts

\usepackage{cite}
\usepackage{url}
\usepackage{fancyvrb}
\usepackage[final]{pdfpages}
\usepackage{longtable}
\usepackage{booktabs}
\usepackage{amsmath}
\usepackage{amssymb}
\usepackage[linesnumbered, ruled, boxed]{algorithm2e} 
\usepackage{makecell}
\usepackage{soul}
\usepackage{graphicx}

\ifCLASSINFOpdf
\else
\fi

\begin{document}
\newtheorem{theorem}{Theorem}
\newtheorem{acknowledgement}[theorem]{Acknowledgement}
\newtheorem{axiom}[theorem]{Axiom}
\newtheorem{case}[theorem]{Case}
\newtheorem{claim}[theorem]{Claim}
\newtheorem{conclusion}[theorem]{Conclusion}
\newtheorem{condition}[theorem]{Condition}
\newtheorem{conjecture}[theorem]{Conjecture}
\newtheorem{criterion}[theorem]{Criterion}
\newtheorem{definition}{Definition}
\newtheorem{exercise}[theorem]{Exercise}
\newtheorem{lemma}{Lemma}
\newtheorem{corollary}{Corollary}
\newtheorem{notation}[theorem]{Notation}
\newtheorem{problem}[theorem]{Problem}
\newtheorem{proposition}{Proposition}
\newtheorem{scheme}{Scheme}   
\newtheorem{solution}[theorem]{Solution}
\newtheorem{summary}[theorem]{Summary}
\newtheorem{assumption}{Assumption}
\newtheorem{example}{\bf Example}
\newtheorem{remark}{\bf Remark}

\def\qed{$\Box$}
\def\QED{\mbox{\phantom{m}}\nolinebreak\hfill$\,\Box$}
\def\proof{\noindent{\emph{Proof:} }}
\def\poof{\noindent{\emph{Sketch of Proof:} }}
\def
\endproof{\hspace*{\fill}~\qed
\par
\endtrivlist\unskip}
\def\endproof{\hspace*{\fill}~\qed\par\endtrivlist\vskip3pt}

\def\E{\mathsf{E}}
\def\eps{\varepsilon}
\def\phi{\varphi}
\def\Lsp{{\boldsymbol L}}
\def\Bsp{{\boldsymbol B}}
\def\lsp{{\boldsymbol\ell}}
\def\Ltsp{{\Lsp^2}}
\def\Lpsp{{\Lsp^p}}
\def\Linsp{{\Lsp^{\infty}}}
\def\LtR{{\Lsp^2(\Rst)}}
\def\ltZ{{\lsp^2(\Zst)}}
\def\ltsp{{\lsp^2}}
\def\ltZt{{\lsp^2(\Zst^{2})}}
\def\ninN{{n{\in}\Nst}}
\def\oh{{\frac{1}{2}}}
\def\grass{{\cal G}}
\def\ord{{\cal O}}
\def\dist{{d_G}}
\def\conj#1{{\overline#1}}
\def\ntoinf{{n \rightarrow \infty }}
\def\toinf{{\rightarrow \infty }}
\def\tozero{{\rightarrow 0 }}
\def\trace{{\operatorname{Tr}}}
\def\ord{{\cal O}}
\def\UU{{\cal U}}
\def\rank{{\operatorname{rank}}}
\def\acos{{\operatorname{acos}}}

\def\SINR{\mathsf{SINR}}
\def\SNR{\mathsf{SNR}}
\def\SIR{\mathsf{SIR}}
\def\tSIR{\widetilde{\mathsf{SIR}}}
\def\Ei{\mathsf{Ei}}
\def\l{\left}
\def\r{\right}
\def\({\left(}
\def\){\right)}
\def\lb{\left\{}
\def\rb{\right\}}

\setcounter{page}{1}

\newcommand{\eref}[1]{(\ref{#1})}
\newcommand{\fig}[1]{Fig.\ \ref{#1}}

\def\bydef{:=}
\def\ba{{\mathbf{a}}}
\def\bb{{\mathbf{b}}}
\def\bc{{\mathbf{c}}}
\def\bd{{\mathbf{d}}}
\def\bee{{\mathbf{e}}}
\def\bff{{\mathbf{f}}}
\def\bg{{\mathbf{g}}}
\def\bh{{\mathbf{h}}}
\def\bi{{\mathbf{i}}}
\def\bj{{\mathbf{j}}}
\def\bk{{\mathbf{k}}}
\def\bl{{\mathbf{l}}}
\def\bn{{\mathbf{n}}}
\def\bo{{\mathbf{o}}}
\def\bp{{\mathbf{p}}}
\def\bq{{\mathbf{q}}}
\def\br{{\mathbf{r}}}
\def\bs{{\mathbf{s}}}
\def\bt{{\mathbf{t}}}
\def\bu{{\mathbf{u}}}
\def\bv{{\mathbf{v}}}
\def\bw{{\mathbf{w}}}
\def\bx{{\mathbf{x}}}
\def\by{{\mathbf{y}}}
\def\bz{{\mathbf{z}}}
\def\b0{{\mathbf{0}}}

\def\bA{{\mathbf{A}}}
\def\bB{{\mathbf{B}}}
\def\bC{{\mathbf{C}}}
\def\bD{{\mathbf{D}}}
\def\bE{{\mathbf{E}}}
\def\bF{{\mathbf{F}}}
\def\bG{{\mathbf{G}}}
\def\bH{{\mathbf{H}}}
\def\bI{{\mathbf{I}}}
\def\bJ{{\mathbf{J}}}
\def\bK{{\mathbf{K}}}
\def\bL{{\mathbf{L}}}
\def\bM{{\mathbf{M}}}
\def\bN{{\mathbf{N}}}
\def\bO{{\mathbf{O}}}
\def\bP{{\mathbf{P}}}
\def\bQ{{\mathbf{Q}}}
\def\bR{{\mathbf{R}}}
\def\bS{{\mathbf{S}}}
\def\bT{{\mathbf{T}}}
\def\bU{{\mathbf{U}}}
\def\bV{{\mathbf{V}}}
\def\bW{{\mathbf{W}}}
\def\bX{{\mathbf{X}}}
\def\bY{{\mathbf{Y}}}
\def\bZ{{\mathbf{Z}}}

\def\mA{{\mathbb{A}}}
\def\mB{{\mathbb{B}}}
\def\mC{{\mathbb{C}}}
\def\mD{{\mathbb{D}}}
\def\mE{{\mathbb{E}}}
\def\mF{{\mathbb{F}}}
\def\mG{{\mathbb{G}}}
\def\mH{{\mathbb{H}}}
\def\mI{{\mathbb{I}}}
\def\mJ{{\mathbb{J}}}
\def\mK{{\mathbb{K}}}
\def\mL{{\mathbb{L}}}
\def\mM{{\mathbb{M}}}
\def\mN{{\mathbb{N}}}
\def\mO{{\mathbb{O}}}
\def\mP{{\mathbb{P}}}
\def\mQ{{\mathbb{Q}}}
\def\mR{{\mathbb{R}}}
\def\mS{{\mathbb{S}}}
\def\mT{{\mathbb{T}}}
\def\mU{{\mathbb{U}}}
\def\mV{{\mathbb{V}}}
\def\mW{{\mathbb{W}}}
\def\mX{{\mathbb{X}}}
\def\mY{{\mathbb{Y}}}
\def\mZ{{\mathbb{Z}}}

\def\cA{\mathcal{A}}
\def\cB{\mathcal{B}}
\def\cC{\mathcal{C}}
\def\cD{\mathcal{D}}
\def\cE{\mathcal{E}}
\def\cF{\mathcal{F}}
\def\cG{\mathcal{G}}
\def\cH{\mathcal{H}}
\def\cI{\mathcal{I}}
\def\cJ{\mathcal{J}}
\def\cK{\mathcal{K}}
\def\cL{\mathcal{L}}
\def\cM{\mathcal{M}}
\def\cN{\mathcal{N}}
\def\cO{\mathcal{O}}
\def\cP{\mathcal{P}}
\def\cQ{\mathcal{Q}}
\def\cR{\mathcal{R}}
\def\cS{\mathcal{S}}
\def\cT{\mathcal{T}}
\def\cU{\mathcal{U}}
\def\cV{\mathcal{V}}
\def\cW{\mathcal{W}}
\def\cX{\mathcal{X}}
\def\cY{\mathcal{Y}}
\def\cZ{\mathcal{Z}}
\def\cd{\mathcal{d}}
\def\Mt{M_{t}}
\def\Mr{M_{r}}
\def\O{\Omega_{M_{t}}}
\newcommand{\figref}[1]{{Fig.}~\ref{#1}}
\newcommand{\tabref}[1]{{Table}~\ref{#1}}

\newcommand{\var}{\mathsf{var}}
\newcommand{\fb}{\tx{fb}}
\newcommand{\nf}{\tx{nf}}
\newcommand{\BC}{\tx{(bc)}}
\newcommand{\MAC}{\tx{(mac)}}
\newcommand{\Pout}{p_{\mathsf{out}}}
\newcommand{\nnn}{\nn\\}
\newcommand{\FB}{\tx{FB}}
\newcommand{\TX}{\tx{TX}}
\newcommand{\RX}{\tx{RX}}
\renewcommand{\mod}{\tx{mod}}
\newcommand{\m}[1]{\mathbf{#1}}
\newcommand{\td}[1]{\tilde{#1}}
\newcommand{\sbf}[1]{\scriptsize{\textbf{#1}}}
\newcommand{\stxt}[1]{\scriptsize{\textrm{#1}}}
\newcommand{\suml}[2]{\sum\limits_{#1}^{#2}}
\newcommand{\sumlk}{\sum\limits_{k=0}^{K-1}}
\newcommand{\eqhsp}{\hspace{10 pt}}
\newcommand{\tx}[1]{\texttt{#1}}
\newcommand{\Hz}{\ \tx{Hz}}
\newcommand{\sinc}{\tx{sinc}}
\newcommand{\tr}{\mathrm{tr}}
\newcommand{\diag}{\mathrm{diag}}
\newcommand{\MAI}{\tx{MAI}}
\newcommand{\ISI}{\tx{ISI}}
\newcommand{\IBI}{\tx{IBI}}
\newcommand{\CN}{\tx{CN}}
\newcommand{\CP}{\tx{CP}}
\newcommand{\ZP}{\tx{ZP}}
\newcommand{\ZF}{\tx{ZF}}
\newcommand{\SP}{\tx{SP}}
\newcommand{\MMSE}{\tx{MMSE}}
\newcommand{\MINF}{\tx{MINF}}
\newcommand{\RC}{\tx{MP}}
\newcommand{\MBER}{\tx{MBER}}
\newcommand{\MSNR}{\tx{MSNR}}
\newcommand{\MCAP}{\tx{MCAP}}
\newcommand{\vol}{\tx{vol}}
\newcommand{\ah}{\hat{g}}
\newcommand{\tg}{\tilde{g}}
\newcommand{\teta}{\tilde{\eta}}
\newcommand{\heta}{\hat{\eta}}
\newcommand{\uh}{\m{\hat{s}}}
\newcommand{\eh}{\m{\hat{\eta}}}
\newcommand{\hv}{\m{h}}
\newcommand{\hh}{\m{\hat{h}}}
\newcommand{\Po}{P_{\mathrm{out}}}
\newcommand{\Poh}{\hat{P}_{\mathrm{out}}}
\newcommand{\Ph}{\hat{\gamma}}
\newcommand{\mat}[1]{\begin{matrix}#1\end{matrix}}
\newcommand{\ud}{^{\dagger}}
\newcommand{\C}{\mathcal{C}}
\newcommand{\nn}{\nonumber}
\newcommand{\nInf}{U\rightarrow \infty}

\title{\huge Channel Capacity-Aware Distributed Encoding for\\ Multi-View Sensing and Edge Inference}

\author{ \IEEEauthorblockN{Mingjie Yang\IEEEauthorrefmark{1}, Guangming Liang\IEEEauthorrefmark{1}, Dongzhu Liu\IEEEauthorrefmark{1}, Lei Zhang\IEEEauthorrefmark{2}, and Kaibin Huang \IEEEauthorrefmark{3} 
}
\IEEEauthorblockA{\IEEEauthorrefmark{1}{School of Computing Science, University of Glasgow} \\
\IEEEauthorrefmark{2}{James Watt School of Engineering, University of Glasgow} \\
\IEEEauthorrefmark{3}{Department of Electrical and Electronic Engineering, The University of Hong Kong} \\ 
Email: 
\{2921021y, 3032221l\}@student.gla.ac.uk,     \{dongzhu.liu,  lei.zhang\}@glasgow.ac.uk, huangkb@eee.hku.hk
    }
}

\maketitle

\begin{abstract}

Integrated sensing and communication (ISAC) unifies wireless communication and sensing by sharing spectrum and hardware, which often incurs trade-offs between two functions due to limited resources. However, this paper shifts focus to exploring the synergy between communication and sensing, using WiFi sensing as an exemplary scenario where communication signals are repurposed to probe the environment without dedicated sensing waveforms, followed by data uploading to the edge server for inference. While increased device participation enhances multi-view sensing data, it also imposes significant communication overhead between devices and the edge server. 
To address this challenge, we aim to maximize the sensing task performance,  measured by mutual information, under the channel capacity constraint. The information-theoretic optimization problem is solved by the proposed ADE-MI, a novel framework that employs a two-stage optimization two-stage optimization approach: (1) adaptive distributed encoding (ADE) at the device, which ensures transmitted bits are most relevant to sensing tasks, and (2) multi-view Inference (MI) at the edge server, which orchestrates multi-view data from distributed devices. Our experimental results highlight the synergy between communication and sensing, showing that more frequent communication from WiFi access points to edge devices improves sensing inference accuracy. The proposed ADE-MI  achieves 92\% recognition accuracy with over $10^4$-fold reduction in latency compared to schemes with raw data communication, achieving both high sensing inference accuracy and low communication latency simultaneously.

\end{abstract}
\begin{IEEEkeywords}
    Integrated Sensing and Communication, Task-Oriented Communication, Multi-View Learning,  WiFi Sensing 
\end{IEEEkeywords}

%
\IEEEpeerreviewmaketitle

\vspace{-1mm}
\section{Introduction}


Integrated sensing and communication (ISAC) represents a revolutionary paradigm shift in wireless systems that seamlessly combines sensing and communication functionalities through shared resources, thereby achieving significant improvements in spectral and energy efficiency \cite{9737357}. However, treating communication and sensing as independent functions often leads to performance trade-offs under limited resource constraints. This challenge has motivated the development of innovative resource allocation strategies, such as beamforming design \cite{liu2020radar}, power allocation \cite{li2022novel}, and time scheduling \cite{9933849}.
Although these approaches advance resource management, their focus on resolving resource competition between communication and sensing overlooks the potential for a synergistic integration of the two functionalities.

Among various ISAC implementations, WiFi sensing naturally embodies the dual-functionality concept by leveraging wireless signals for both communication and sensing purposes without requiring dedicated sensing waveforms~\cite{he2023sencom}.  While existing research has successfully demonstrated the feasibility of WiFi-based sensing across various applications, 
they mainly focus on enhancing sensing capabilities through sophisticated signal processing techniques, e.g., by filtering and feature extraction to capture the periodic motion associated with breathing encapsulating in channel state information~(CSI)~\cite{wang2016human}, by using dynamic fresnel zone model to recognize activity-related variations in the CSI~\cite{liu2023towards} and time-frequency analysis to identify Doppler shifts caused by gesture movements~\cite{9516988}. However, their efforts predominantly focus on improving sensing performance without exploring the interplay between sensing and communication.

The sensing data collected in ISAC systems supports downstream inference tasks, which serve as performance indicators for the sensing process. 
This introduces intensive computational load to the ISAC paradigm, thereby giving rise to the integrated sensing, communication, and computation (ISCC) paradigm~\cite{10293204,wen2024survey}. Inference can be performed either locally on the device or offloaded to an edge server. On-device inference processes are often constrained by limited computational resources \cite{zalewski2020bits}. With the rise of edge AI, edge computing infrastructure provides an alternative by offloading inference tasks to edge servers to achieve a better inference performance~\cite{10574240}. 
While this approach enriches data from distributed devices for collaborative inference, it also introduces significant communication overhead due to data uploading, especially when more devices participate in sensing. To reduce communication costs, edge devices often apply  principal component analysis (PCA) for dimensionality reduction, followed by techniques such as over-the-air computation \cite{10230049} or power allocation \cite{10669354} to further enhance communication efficiency. However, these methods fail to fully align the sensing function — defined here as inference performance — with the communication function, as PCA cannot guarantee that the extracted features are optimal for the inference tasks.

This paper investigates multi-view WiFi sensing and edge inference as a representative scenario to demonstrate the synergy of communication and sensing. As motivated by task-oriented communication \cite{8970161} and multi-device cooperative edge inference \cite{9837474}, we aim to transmit the most informative features to facilitate sensing performance under the channel capacity constraint. 
Unlike \cite{9837474}, which overlooks practical wireless channel conditions, our approach incorporates channel capacity awareness into the encoding process, enabling a more effective synergy between sensing and communication.

We formulate the problem using an information-theoretic approach and approximate the solution via variational distributions and Monte Carlo sampling. Since the problem involves joint learning of the local encoder and the inference model at the edge server, frequent communication is required during the training phase. To address this, we decompose the original problem into two stages of optimization: 1) adaptive distributed encoding (ADE), which dynamically adjusts the dimensionality of encoded sensing data based on channel capacity, and 2) multi-view inference (MI) at the edge server, yielding ADE-MI that only requires one-shot communication. Our experimental results demonstrate that more frequent communication from WiFi access points to edge devices improves sensing inference accuracy, and the proposed ADE-MI framework achieves a 92\% recognition accuracy while reducing latency by over $10^4$-fold compared to raw data transmission, achieving both high sensing inference accuracy and low communication latency.

\vspace{-2mm}
\section{System Model}\label{Sec2}
\begin{figure}[t]
    \centering 
    \includegraphics[scale=0.37]{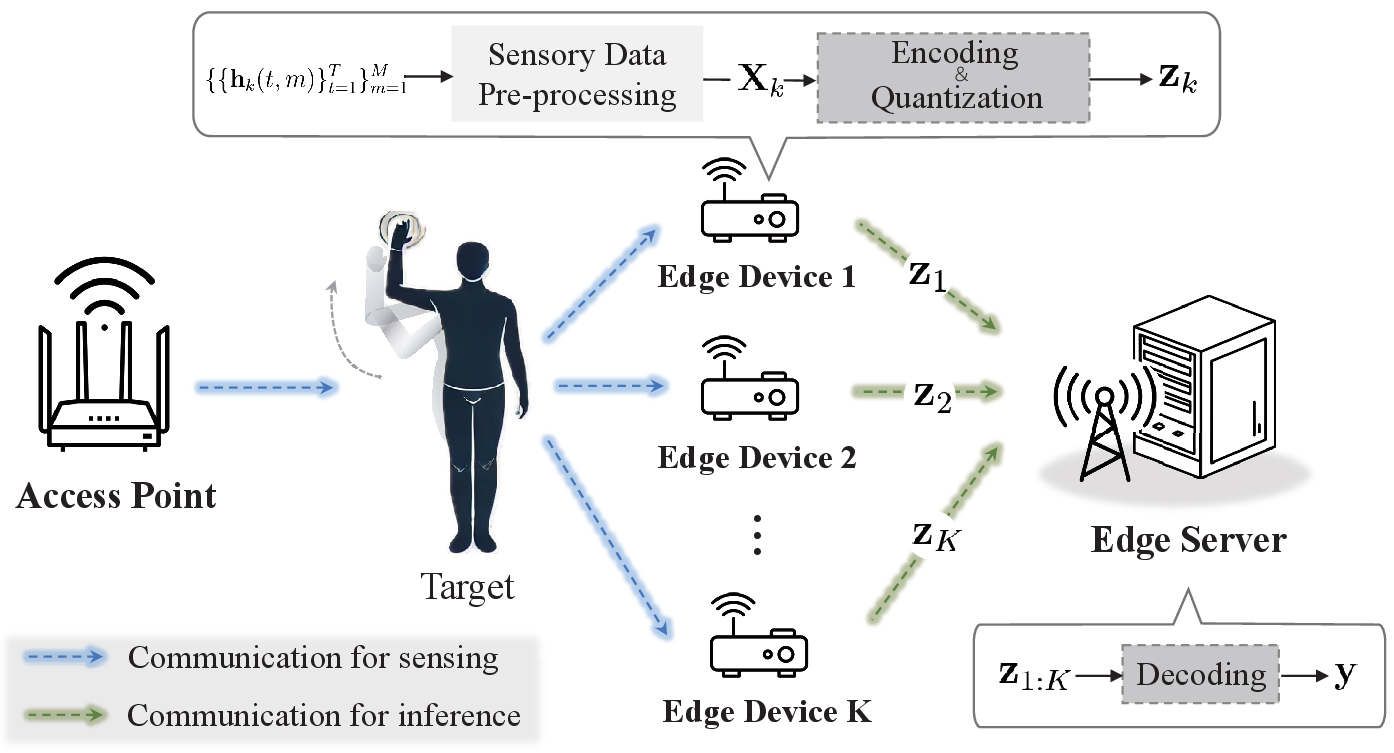}
    \caption{Overview of the adaptive distributed encoding-based multi-view gesture recognition system.}
    \label{fig:system_model}
\end{figure}

We consider a multi-view WiFi sensing system as shown in Fig.~\ref{fig:system_model}.  At a regular communication interval $\Delta t$, the access point (AP) simultaneously communicates with $K$ edge devices via orthogonal frequency-division multiple access (OFDMA) channel.  The communication signals are dual used for transmitting data streams and measuring CSI during $T$ 
seconds for sensing. 
Each WiFi device processes the received signals to extract low-dimensional features, which are subsequently uploaded to the edge server for inference the human action. The AP is equipped with $N$ antennas, while each WiFi device and the edge server are equipped with a single antenna.

\vspace{-2mm}

\subsection{Sensing Signal Pre-Processing}

In WiFi sensing, each device uses the preamble, a critical part of the WiFi packet, for channel estimation. At time slot $t$ and subcarrier $m$, the extracted CSI at device $k$, denoted as $\mathbf{h}_{k}(t,m)\in \mathbb{C}^{N\times 1}$  is represented as follows
\vspace{-1mm}
\begin{equation}\label{eq: channel} 
    \mathbf{h}_{k}(t,m) \!\!= \!\! \left[\sum_{l=1}^L \boldsymbol{\alpha}_{k,l}(t,m) \circ e^{j[\boldsymbol{\phi}_{k,l}(t,m)+2 \pi {\bf f}_{k,l}^D(t,m)]}\right]e^{j\epsilon(t,m)},
\end{equation} 
where $\circ$ denotes the element-wise product, $l=1,\cdots, L$ is the multipath index, $\boldsymbol{\alpha}_{k,l}(t,m)\in \mathbb{R}^d$ is the channel gain vector,   $\boldsymbol{\phi}_{k,l}(t,m) \in \mathbb{R}^d$ is the phase shift due to time delay, and ${\bf f}_{k,l}^D(t,m)\in \mathbb{R}^d$ is the Doppler frequency shift (DFS), and $\epsilon(t,m) \in \mathbb{R} $ is the phase error incurred by timing alignment offset.  
To extract DFS, one antenna is selected as the reference, and conjugate multiplication is performed with the other $N-1$ antennas to eliminate the phase error~\cite{wu2022wifi}. This is followed by applying the low-pass filter to remove the quasi-static channel components and the high-pass filter to remove the high-frequency noise. The processed CSI is represented as a matrix $\mathbf{F}_k \in \mathbb{C}^{S \times (N-1)M}$, where $S=T/\Delta t$ is the number of samples over time, and $M$ is the number of subcarriers allocated to each device. 
$\mathbf{F}_k$ predominantly reflects motion-induced frequency changes. 

Since the subcarriers and antennas in CSI are correlated~\cite{ali2015keystroke}, and each subcarrier captures different center frequencies and DFS associated with the wavelength of the signal, principal component analysis (PCA) is applied to extract the first dominant component of $\mathbf{F}_k $ as 
\begin{align}
\mathbf{c}_k= \mathbf{F}_k \mathbf{V}_{k}(:,1), 
\end{align}
where we have $\mathbf{F}_k=\mathbf{U}_k\mathbf{\Sigma}_k\mathbf{V}_k^{H} $, and $\mathbf{V}_{k}(:,1)$ is the first principal vector of $\mathbf{V}_k$. 
Further applying short-time Fourier transform~(STFT) to $\mathbf{c}_k$  yields the spectrogram $\mathbf{X}_k \in \mathbb{R}^{S_{T} \times S_F} $, which captures the power distribution across both time and Doppler frequency domains~\cite{ma2019wifi}. Here, $S_T $ is the number of STFT frames, calculated as $S_T = \lceil\frac{S-W}{L_{hop}}\rceil$ where $W$ and $L_{hop}$ are the window length and the window hop size, respectively. $S_F$  is the number of frequency bins, corresponding to the frequency resolution of the STFT. The spectrogram  $\mathbf{X}_k$  serves as a DFS profile that can be directly used for human action recognition. However, transmitting the high-dimensional time-frequency DFS profile  $\mathbf{X}_k $ to the edge server introduces heavy communication overhead for training and incurs unaffordable latency for real-time edge inference.

\subsection{Probabilistic Learning Model} 
We introduce a probabilistic learning model for multi-view cooperative edge inference. The DFS profiles collected by the edge devices,  ${\bf X}_1, \ldots, {\bf X}_K$, along with the target variable ${\bf y}$, i.e., the label of human actions, are modeled as realizations of the random variables $X_1, \ldots, X_K$ and $Y$, respectively. To reduce communication overhead, each edge device encodes its DFS profile ${\bf X}_k$ into a low-dimensional feature representation ${\bf z}_k \in \mathbb{R}^{d_k}$, which is then transmitted to the edge server. The transmitted feature ${\bf z}_k$  represent realizations of the random variable $Z_k$.
Upon receiving the encoded features from all edge devices, the edge server performs further processing to generate the inference label. The random variables involved follow the Markov chain:
\begin{align}
    Y \leftrightarrow X_k \leftrightarrow Z_k, \quad \forall k \in \{1, \ldots, K\}, 
\end{align}
which satisfies $p({\bf z}_k,{\bf X}_k|{\bf y})=p({\bf z}_k|{\bf X}_k)p({\bf X}_k|{\bf y})$, for all $k\in\{1,\cdots,K\}$. 
The learning model consists of local encoder $p_{\theta_k}({\bf z}_k | {\bf X}_k)$, parameterized by ${\theta_k}$, at each edge device $k$, and the inference model at the edge server parameterized by $\psi$, which models the distribution $p_{\psi}({\bf y} | {\bf z}_1, \ldots, {\bf z}_K)$.
The design target is to maximize the inference performance under the communication constraint, which are elaborated in the following section.

\subsection{Communication Model for Edge Learning}
Each device transmits the low-dimensional quantization vector ${\bf z}_k$ to the edge server via the OFDMA channel. Let $B_k$ denote the bandwidth allocated to device $k$ and $\gamma_k$ its signal-to-noise ratio~(SNR). Then, the maximum transmission rate, denoted by $C_k$, is given by Shannon’s capacity formula: 
\begin{equation}
    C_k = B_k \textup{log}_2(1+\gamma_k).\label{eq:shannon}
\end{equation} 

To achieve reliable communication with an arbitrarily low probability of error, the encoding must satisfy the channel capacity constraint, defined by the mutual information between ${X}_k$ and ${Z}_k$:
 \begin{equation}\label{eq: channel capacity constriant}
     \mathrm{I}({ X}_k;{Z}_k)\leq C_k.  
 \end{equation}
 
The edge server aggregates the received signals to construct a comprehensive DFS representation ${\bf Z}=\{{\bf z}_1,{\bf z}_2,...,{\bf z}_K\}$, enabling multi-view learning to enhance inference accuracy.

\section{Adaptive Distributed Encoding for Multi-View Inference}\label{Sec3}

This section starts with formulating an information-theoretical optimization problem to maximize the sensing inference performance under channel capacity constraint in \eqref{eq: channel capacity constriant}. We adopt the variational approximation and Monte Carlo sampling to address the intractable distributions involved in the optimization problem. To save the communication overhead during training, the problem is divided into two stages of optimization: 1) adaptive local encoding (ADE), which extracts task-relevant features on edge devices, and 2) multi-view inference (MI) by leveraging the aggregated multi-view data on the edge server. The solution approach is summarized in algorithms.

\subsection{Problem Formulation}

Our objective is to maximize the inference accuracy at the edge server under the communication constraint specified by \eqref{eq: channel capacity constriant}. To achieve this, the encoded feature variables $Z_1,\cdots,Z_K$ should jointly capture as much information about the target variable $Y$ as possible. This objective can be quantified by the mutual information $\mathrm{I}(Y;Z)$, where $Z=\{Z_1,\cdots, Z_k\}$. 

Maximizing the mutual information  $\mathrm{I}(Y;Z)$ is equivalent to minimizing $\mathrm{H}(Y|Z)$, due to the relationship  $\mathrm{I}(Y;Z)=\mathrm{H}(Y)- \mathrm{H}(Y|Z)$, where $\mathrm{H}(Y)$ is an constant.  Therefore, the problem of multi-view collaborative inference under the communication constraint can be formulated as follows: 
\begin{equation}
\textbf{(P1)} \quad
\begin{aligned} \label{eq:opt1} 
    &\min \quad \mathrm{H}(Y|Z_1, Z_2,\cdots, Z_K) \\
     &\mathrm{s.t.}  \quad \mathrm{I}({ X}_k;{Z}_k)\leq C_k, \quad \forall k\in \{1,2,..,K\}. 
\end{aligned}
\end{equation}

Note that one can introduce Lagrange multiplier $\beta_k \geq 0$ to incorporate the mutual information constraint of each device into the objective function. This converts the original constrained problem into an unconstrained Lagrangian form:
\vspace{-1mm}
\begin{equation}
    \mathcal{L} = \mathrm{H}(Y|{Z}_1,\cdots,Z_K) + \sum_{k=1}^{K} \beta_k \left[\mathrm{I}(X_k;{Z}_k)-C_k\right].
\end{equation}
\vspace{-1mm}
It is the classic information bottleneck principle, where $\beta_k$ controls the balance between maximizing the task-relevant information for prediction accuracy and minimizing redundancy. Although mutual information has broad applications in representation learning, it does not guarantee that the encoded data $\bz_k$ will fit within the bit rate, which is the focus of wireless transmissions. In the next section, we will discuss how to adapt the dimensionality of $\bz_k$ to meet the bit stream constraint in practice.

\subsection{Channel Capacity-Aware Dimensionality Reduction}
To address the communication constraint in terms of bit rate, the local encoder $\theta_k$ at device $k$ consists of a trainable feature extractor followed by a fixed uniform quantization module. In this work, we adopt uniform quantization due to its simplicity in calculating the bit length. 
 The channel capacity $C_k$ regulates the maximum allowable bit-rate for device $k$, which provides critical guidance for encoder design.
 Specifically, given the channel capacity constraint in \eqref{eq:shannon},  the dimensionality of the encoded feature vector $\bz_k$, denoted by $d_k$, is determined as
\begin{equation}
d_k = \left \lfloor \frac{C_k}{n_k L}\right \rfloor, 
\label{eq:dim}
\end{equation}
where $n_k$ is the number of bits used for quantizing each element, and $L$ is the number of training samples uploaded per transmission.

\subsection{Variational Approximation for Distributed Encoding and Multi-View Inference}
The remaining of this section focus on how to approximate the objective function via variational distribution to enable distributed encoding and multi-view inference. 
Recall that we have defined the local encoder for the distribution $p_{\theta_k}({\bf z}_k | {\bf X}_k)$ and the inference model at the edge server for distribution $p_{\psi}({\bf y} | {\bf z}_1, \ldots, {\bf z}_K)$. The conditional entropy in \eqref{eq:opt1} can be rewritten as 
\vspace{-2.5mm}
\begin{align}
    \mathrm{H}(Y|Z_1,\cdots,Z_K) =  - \E_{p(\by, \bz_1,\cdots, \bz_K)}\left[\log p_{\psi}({\bf y} | {\bf z}_1, \ldots, {\bf z}_K) \right], \label{eq:con_entropy}
\end{align}
Given the Markov chain $Y \leftrightarrow X_k \leftrightarrow Z_k$,  the joint distribution $p(\by, \bz_1,\cdots, \bz_K)$ can be factorized as
\vspace{-2mm}
\begin{align}\label{eq: joint dis}
&p(\by, \mathbf{z}_1, \ldots, \mathbf{z}_K) \nn \\ 
&= p(\by) \int  \prod_{k=1}^K p(\bX_k|\by) p_{\theta_k}(\mathbf{z}_k \mid \mathbf{X}_k) \, d\mathbf{X}_1 \cdots d\mathbf{X}_K. 
\end{align}
Substituting \eqref{eq: joint dis} into \eqref{eq:con_entropy}, we have 
\vspace{-1.5mm}
\begin{align}
    &\mathrm{H}(Y|Z_1,\cdots, Z_K)\nn\\ 
    &=  - \int p(\by) \int \int \prod_{k=1}^K  p(\mathbf{X}_k \mid \by) p_{\theta_k}(\mathbf{z}_k \mid \mathbf{X}_k)  \nn \\ 
    \vspace{-1mm}
    & \qquad  \qquad \qquad  \times \log p_\psi(\by \mid \mathbf{z}_1, \ldots, \mathbf{z}_K)  d \bX_{1:k}   d\bz_{1:K} \, d\by  \nn \\ 
    \vspace{-1mm}
    &\approx -\frac{1}{L}\sum_{l=1}^L  \log p_{\psi}({\by}^{(l)}|\bz_1^{(l)},\cdots, \bz_K^{(l)}), \label{eq: obj_sampling}
\end{align} 
where the final step uses Monte Carlo approximation with $L$ samples. Specifically, given a data sample $\bX_k^{(l)}$ drawn from $ p(\bX_k|\by^{(l)})$, we sample encoding vector $\bz_k^{(l)}$ from the distribution $ p_{\theta_k}(\bz_k|\bX_k^{(l)})$. This process is repeated for $l=1,\cdots, L$, and the resulting samples are used to approximate the expectation integral in the objective function.

Minimizing \eqref{eq: obj_sampling} requires joint updates $\{\theta_k\}_{k=1}^K$ and $\psi$ across all devices and the edge server. This process involves frequent communication between the server and devices to exchange updates of the encoding vector $\{\bz_1^{(l)}, \cdots, \bz_K^{(l)}\}_{l=1}^L$ and synchronize parameters during each iteration. Such an approach is communication-intensive and poses significant challenges for distributed systems. To address this, we propose decomposing the optimization problem {\bf P1} in \eqref{eq:opt1} into two subproblems: 
\vspace{-2.7mm}
\begin{equation}
\textbf{(P1.1)} \quad
\begin{aligned} \label{eq:opt1-1} 
    &\min \quad \mathrm{H}(Y| Z_k) \\
     &\mathrm{s.t.}  \quad \mathrm{I}({X}_k;{Z}_k)\leq C_k. 
\end{aligned}
\end{equation}
which can be solved independently on each device $k$. Once solved, each device uses its well-trained encoder to generate the encoding vectors $\{\bz_k^{(l)}\}_{l=1}^L$, and uploads them to the edge server in a one-shot transmission. The edge server then solves the multiview inference problem:
\vspace{-1.mm}
\begin{equation}
\textbf{(P1.2)} \quad
\begin{aligned} \label{eq:opt1-2} 
    &\min \quad \mathrm{H}(Y|Z_1,\cdots, Z_K) 
\end{aligned}
\end{equation} 
using the received encoding vectors $\{\bz_1^{(l)}, \cdots, \bz_k^{(l)}\}_{l=1}^L$ and target labels $\{\by^{(l)}\}_{l=1}^L$.  

This decomposition enables a two-stage optimization process, termed adaptive distributed encoding and multi-view inference (ADE-MI). In the first stage, each device  independently trains a local encoder to extract task-relevant features while satisfying the individual channel capacity constraints. In the second stage, the server trains a joint inference model to maximize prediction accuracy by using the encoded multi-view data received from the distributed devices.

\subsubsection{Adaptive Distributed Encoding (ADE)} The optimization problem in {\bf P1.1} is equivalent to minimizing an upper bound of {\bf P1}, as we have 
$\mathrm{H}(Y|Z_1,\cdots, Z_K)\leq \min_k \mathrm{H}(Y|Z_k) $, given the Markov chain  $Y \leftrightarrow X_k \leftrightarrow Z_k$. 

Following a similar approach to that in \eqref{eq: obj_sampling}, we employ an auxiliary local inference model 
$\phi_k$ to approximate the distribution as 
\vspace{-1mm}
\begin{align}
p_{\phi_k }(\by |\bz_k) \propto \exp(-\ell(\by, \hat{\by}(\mathbf{z}_k; \phi_k))),  \label{eq: local inference}
\end{align}
where $\ell(\cdot,\cdot) $ denotes the loss between the predicted value $\hat{\by}$ and the ground truth $\by$. Let $\tilde{\bz}_k$ represent the unquantized version of ${\bz}_k$.  We model its distribution $p_{\theta_k} ( \tilde{\bz}_k|\bX_k)$ as a multivariate Gaussian with mean $\boldsymbol{\mu}(\bX_k; {\theta_k}) \in \mathbb{R}^{d_k}$ and and diagonal covariance $ \boldsymbol{\sigma}(\bX_k; {\theta_k}) \in \mathbb{R}^{d_k}$.  Both $\boldsymbol{\mu}$ and $\boldsymbol{\sigma}$  are outputs of a neural network parameterized by ${\theta_k}$ with input $\bX_k$. Using the reparameterization trick, we sample $\tilde{\bz}_k$ and then obtain $\bz_k$ via quantization as follows:  \vspace{-0.5mm}
\begin{align} 
\tilde{\bz}_k &= \boldsymbol{\mu}(\bX_k; {\theta_k})  + \boldsymbol{\sigma}(\bX_k; {\theta_k})  \odot \epsilon  \label{eq:reparamete}   \\ 
{\bz}_k &= \mathcal{Q}(\tilde{\bz}_k), \label{eq:discrete}
\end{align}
where $\epsilon \sim \mathcal{N}(0,\mathbf{I})$, $\odot$ denotes element-wise product, and $\mathcal{Q}(\cdot)$ represents the uniform quantization. 
Given a batch of data $\{(\bX_k^{(l)}, \by^{(l)})\}_{l=1}^L$, we approximate the objective function $\mathrm{H}(Y| Z_k)$  using Monte Carlo sampling as  
\vspace{-0.5mm}
\begin{equation}
\mathcal{L}_{\text{device},k}(\theta_k,\phi_k) \simeq  - \frac{1}{L}\sum_{l=1}^L \log p_{\phi_k}(\by^{(l)}|\bz_k^{(l)}).\label{eq:loss1}
\end{equation}
The optimization problem {\bf P1.1} can then be solved by minimizing \eqref{eq:loss1} with respect to $\theta_k$ and $\phi_k$ as described in Algorithm~1. The constraint is satisfied by adapting the dimensionality of $\bz_k$ according to  \eqref{eq:dim}. 

\subsubsection{Multi-View Inference (MI)} 
Upon receiving the encoded features $\{\bz_1^{(l)}, \cdots, \bz_K^{(l)}\}_{l=1}^L$  and their  associated labels  $\{\by^{(l)}\}_{l=1}^L$, the distribution $p_{\psi} (\by|\bz_1,\cdots,\bz_K)$ is approximated as in \eqref{eq: local inference}. The inference mode $\psi$ produces the predicted output $\hat{\by}(\bZ;\psi)$ with the input   $\bZ=\{\bz_1, \cdots, \bz_K\}$  represents the multi-view encoded data. 
The objective function in \eqref{eq:opt1-2} is then approximated using Monte Carlo sampling:
\vspace{-0.5mm}
\begin{equation}
\mathcal{L}_{\text{server}}(\psi) \simeq \frac{1}{L} \sum_{l=1}^L -\log p_{\psi}(\by^{(l)}|\bz_1^{(l)},\cdots \bz_K^{(l)}).\label{eq:loss2}
\end{equation} 
The optimization problem {\bf P1.2} is solved by minimizing \eqref{eq:loss2} with respect to $\psi$, as outlined in Algorithm 2.

\enlargethispage{0.1in}
\begin{algorithm}[t]
\caption{On-device Local Training for Adaptive Encoding}
\KwIn{Local training dataset $\mathcal{D}_k$, learning rate $\eta_k$, batch size $L$, number of epochs $I_k$, channel capacity $C_k$}
Compute dimensionality $d_k$ via \eqref{eq:dim};

Initialize parameters $\theta_k^{(0)}$, $\phi_k^{(0)}$ according to $d_k$ \;

\For{$i = 1$ to $I_k$}{
    Randomly select a batch $\{(\bX^{(l)}_{k},\by^{(l)})\}_{l=1}^{L}$ from~$\mathcal{D}_k$\;
  
        \For{$l = 1$ to $L$}{

              Compute $\boldsymbol{\mu}_{\theta_k^{(i-1)}}(\bX^{(l)}_{k})$ and $\boldsymbol{\sigma}_{\theta_k^{(i-1)}}(\bX^{(l)}_{k})$\;

            Sample $\epsilon_k^{(l)} \sim \mathcal{N}(0, \mathbf{I})$\;
            
            Compute $\bz_k^{(l)}$ based on \eqref{eq:reparamete} and \eqref{eq:discrete} \;

        }

        Update parameters $\theta_k$ and $\phi_k$ based on \eqref{eq:loss1} \\
        $\theta_k^{(i)} = \theta_k^{(i-1)} - \eta_k \nabla_{\theta_k}^{(i)}\mathcal{L}_{\text{device},k}(\theta_k^{(i-1)},\phi_k^{(i-1)})$\\
        $\phi_k^{(i)} = \phi_k^{(i-1)} - \eta_k \nabla_{\phi_k}^{(i)}\mathcal{L}_{\text{device},k}(\theta_k^{(i-1)},\phi_k^{(i-1)})$
    }

Given $\theta_k^{(I_k)}$, sample $\bz_k^{(l)}$ for $L$ times based on \eqref{eq:reparamete} and \eqref{eq:discrete}\;

\KwOut{$\{\bz_k^{(l)}\}_{l=1}^L$ and optimized parameters $\theta_k^{(I_k)}$ }
\end{algorithm}
\vspace{-3mm}
\begin{algorithm}[t]
\caption{Multi-view Inference Model Training at the Edge Server}
\KwIn{$\{\bz_1^{(l)},\bz_2^{(l)},\cdots,\bz_K^{(l)}\}_{l=1}^L$ and its corresponding target $\{\by^{l}\}_{l=1}^L$, learning rate $\eta$, number of epochs $I$}

Randomly initialize parameters: $\psi^{(0)}$ \;

\For{$i = 1$ to $I$}{
     Update parameters $\psi$ based on \eqref{eq:loss2} 
    $\psi^{(i)} = \psi^{(i-1)} - \eta\nabla_{\psi}^{(i)}\mathcal{L}_{\text{server}}(\psi^{(i-1)})$
}
\KwOut{Optimized parameters $\psi^{(I)}$}
\end{algorithm}
\vspace{2.2mm}

\section{Experimental Results}\label{Sec4}

In this section, we evaluate the performance of our ADE-MI framework using the Widar3.0 dataset~\cite{9516988} collected in a classroom environment. The sensing area is a $2\text{m}\times 2\text{m}$ square space with an AP placed at one corner, and three WiFi Network Interface Card~(NIC)-equipped devices are strategically set to monitor CSI measurements, providing multiple perspectives for sensing. The dataset contains 9,000 gesture samples across six categories (push-pull, sweep, clap, slide, draw zig-zag, and draw N) labeled from 0 to 5, with 3,000 samples collected from each device. We split the dataset into $90\%$ for training and $10\%$ for testing. 
\subsection{Baseline}
For performance evaluation, we compare ADE-MI against two baselines. The first is single-view sensing utilizing DFS data from only one device without any encoding or compression at the device side and directly transmits it to server. The second baseline is multi-view sensing without encoding at device side, which incorporates raw DFS data from all three devices and transmits it directly to the server. For fair comparison, all methods share identical convolutional neural network~(CNN) structures for feature extraction and deep neural network~(DNN) architectures for gesture inference at the server side. The key distinction lies in that ADE-MI have an additional device-side encoding step before transmission, which addresses practical wireless channel constraints that baselines overlook.
\vspace{-0.9mm}

\subsection{Model Setting}
ADE-MI is implemented in PyTorch, processing input data in the form of preprocessed DFS features with dimensionality of $S_T \times S_F \times K$, where $S_T = 2895$ represents the number of STFT frames, $S_F = 121$ denotes the frequency bins ranging from $-60$ to $+60$ Hz after STFT processing, and $K=3$ is the number of devices. The architecture consists of a pair of encoder and decoder at each device side and a joint decoder at the server side.

The device-specific encoder includes a CNN-based feature extraction module with sequential Conv2D, MaxPool2D, Flatten, and fully connected layers, using ReLU activation. This is followed by DNN-based mean and variance modules ($\boldsymbol{\mu}_{\theta}$ and $\boldsymbol{\sigma}_{\theta}$), each with dimensions of $256\times32$, which output a latent vector with dimension $d_k$. The latent vector is then passed as the input of the paired device-side decoder with a structure of $d_k\times64\times6$ units. At the server side, the joint decoder concatenates the latent vectors from all devices, passing them through a module with $3d_k\times64\times6$ units. The final output layer has 6 units, corresponding to the gesture class predictions for both the device and server models. The training of both local and server-side encoders are both with the same learning rate, $\eta_k = \eta = 1\times10^{-3}$, over $50$ epochs and the batch size of $64$.

\vspace{-2mm}
\subsection{Results}
\begin{figure}[t]
    \centering
    \includegraphics[width=0.3\textwidth]{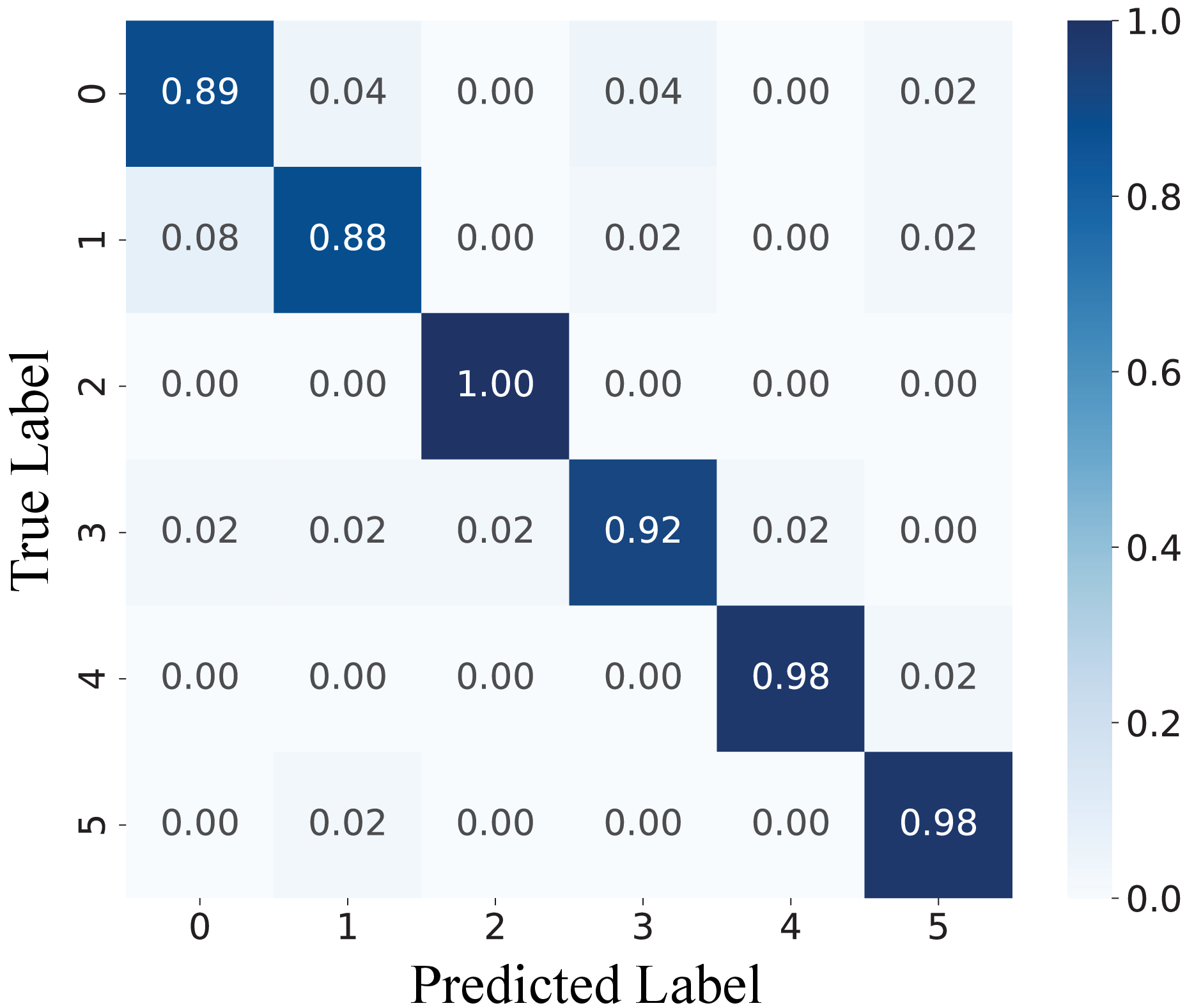}
    \caption{Confusion matrix representing six-gesture classification accuracy for ADE-MI framework.}
    \label{fig:confusion}
\end{figure}

Unless otherwise specified, all experiments are conducted under the following settings: total channel bandwidth $B = 40~\text{MHz}$, SNR $\gamma = 10~\text{dB}$ and available upload time window $100$ s.

Fig.~\ref{fig:confusion} evaluates the classification accuracy of ADE-MI across different gesture types. The results show consistently high recognition rates across all gesture categories, with most gestures achieving accuracy above 90\%. This balanced performance demonstrates the robustness and reliability of our system in practical scenarios.

\begin{figure}[t]
    \centering
    \includegraphics[width=5.5cm]{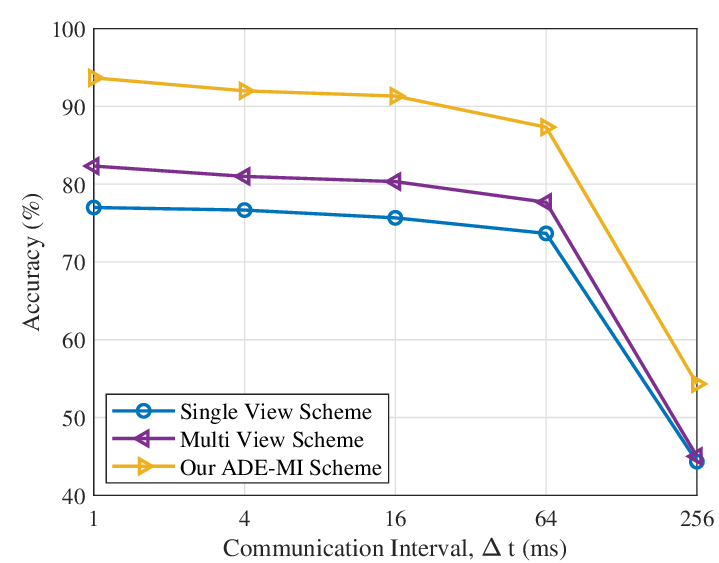}
    \caption{Effect of communication interval on gesture recognition accuracy.}
    \label{fig:interval}
\end{figure}

Fig.~\ref{fig:interval} investigates the impact of communication frequency on recognition performance. The results show that shorter intervals between WiFi packet transmissions lead to higher accuracy, as more frequent sampling captures finer temporal dynamics of gestures. ADE-MI consistently outperforms baselines across different sampling intervals, demonstrating its ability to maintain high accuracy even with varying communication conditions.

\begin{figure}[t]
    \centering
    \includegraphics[width=5.5cm]{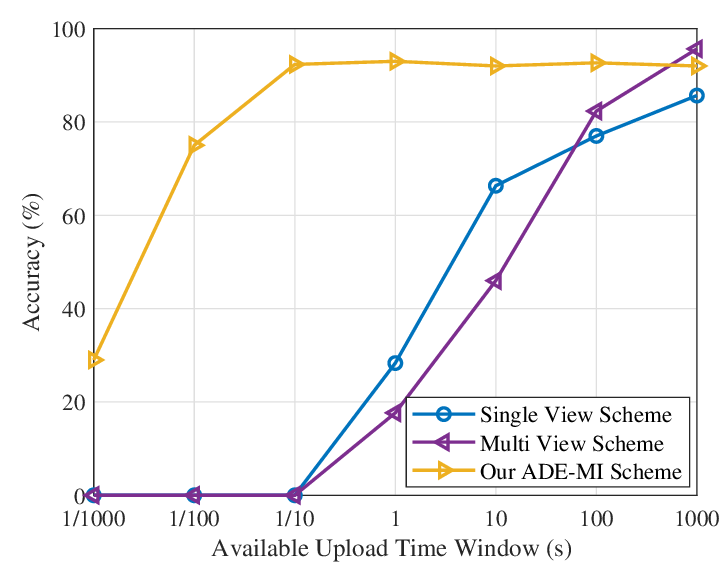}
    \caption{Effect of server communication time on gesture recognition accuracy.}
    \label{fig:latency}
\end{figure}

Fig.~\ref{fig:latency} demonstrates the training efficiency of different approaches. ADE-MI embodies superior convergence speed, reaching 90\% accuracy within just 0.1s of upload time. This dramatic improvement is attributed to dimensionality reduction. More training samples are allowed to be transmitted within bandwidth constraints after being encoded. The results demonstrate that ADE-MI effectively addresses the communication bottleneck in distributed learning while maintaining high accuracy. The rapid convergence and high accuracy highlight the effectiveness of our approach in aligning communication and sensing performance.

As shown in Table~\ref{tab:communication_latency}, we analyze the latency of ADE-MI under varying channel conditions during inference. The results indicate that by dynamically adapting encoder dimensions, our approach maintains consistent low-latency recognition even under challenging SNR scenarios, while baseline methods suffer significant degradation. This adaptability guarantees reliable real-time recognition in dynamic communication environments.

\begin{table}[t]
    \centering
    \caption{Uploading Latency Across SNR Conditions}
    \small
    \setlength\tabcolsep{3pt}
    \begin{tabular}{l *{5}{c}}
        \toprule
        Scheme & 5 dB & 10 dB & 15 dB & 20 dB & 25 dB \\
        \midrule
        Single-View (s)  & 0.27 & 0.16 & 0.11 & 0.08 & 0.07 \\
        Multi-View ~(s)  & 0.80 & 0.48 & 0.33 & 0.25 & 0.20 \\
        ADE-MI$^\dagger$  ($10^{-5} \text{s}$)  & 1.4 & 1.4 & 1.4 & 1.4 & 1.4 \\
        \bottomrule
    \end{tabular}
    \label{tab:communication_latency}
    
    \footnotesize{$^\dagger$Adaptive encoded vector dimensions $d_k$: 6, 10, 15, 19, 24.}
\end{table}
\vspace{-1mm}
\section{Conclusion}\label{Sec5}
This paper presents ADE-MI, a novel multi-view wireless sensing framework that effectively addresses the challenge of reliable gesture recognition under practical wireless channel constraints. By introducing channel-aware encoding at the device side, ADE-MI achieves both high recognition accuracy and low communication overhead. Our experimental results demonstrate that ADE-MI significantly outperforms traditional approaches in terms of latency and adaptability to varying channel conditions, while achieving high recognition accuracy. Future work could explore extending ADE-MI to more diverse application scenarios. Additionally, investigating techniques for further reducing the computational complexity at edge devices while maintaining recognition performance could make the system more energy-efficient.

\bibliographystyle{IEEEtran}
\bibliography{mproj}

\end{document}